\documentstyle[aps,prl]{revtex}
\begin{document}
\author{Sheng Li\thanks{%
lisheng@itp.ac.cn}, Yong Zhang\thanks{%
zhangyo@itp.ac.cn} and Zhongyuan Zhu\thanks{%
zzy@itp.ac.cn}}
\address{$^{*\ddagger }$CCAST World Lab, Academic Sinica, Beijing 100080, P. R. China%
\\
$^{*\dagger \ddagger }$Institute of Theoretical Physics, Academic Sinica, PO%
\\
Box 2735, Beijing 100080, P. R. China}
\title{Decomposition of $SU(N)$ Connection and Effective Theory of $SU(N)$ QCD}
\maketitle

\begin{abstract}
We give a general decomposition of $SU(N)$ connection and derive a
generalized Skyrme-Faddeev action as the effective action of $SU(N)$ QCD in
the low energy limit. The result is obtained by separating the topological
degrees which describes the non-Abelian monopoles from the dynamical degree
of gauge potential, and integrating all the dynamical degrees of $SU(N)$ QCD.
\end{abstract}

\draft

\section{Introduction}

An open question in QCD is to identify the quark confinement mechanism and
understand how it works. The monopole condensation is regarded as a possible
explanation for the confinement of color through the dual Meissner effect%
\cite{th,thooft}. To realize the dual scenario of the confinement, 't Hooft
proposed an Abelian projection where the gauge group is broken by a suitable
gauge condition to its maximal Abelian subgroup\cite{thooft}. Since the
topology of the $SU(N)$ and that of its maximal Abelian subgroup $U(1)^{N-1}$
are different, any such gauge is singular, meaning that a gauge group
element which transforms a generic $SU(N)$ connection onto the gauge fixing
surface is not regular everywhere in space-time. The singularities may form
worldline that are usually interpreted as worldlines of magnetic monopoles.
As a result the original Yang-Mills theory turns into electrodynamics with
magnetic monopoles. Recent numerical simulations show that the monopole
degrees of freedom in the Abelian projection can indeed form a condensation
responsible for the confinement\cite{nu}. However, there still has no
satisfactory proof how the desired monopole condensation could take place in
QCD.

Recently, there are lots of work\cite{cho2,sha} done for searching the
mechanism of confinement in Yang-Mills theory. One hopes to find the
confinement mechanism based on the first principle. Y. M. Cho et al\cite
{cho2} provide a possible theoretical mechanism of the monopole condensation
in $SU(2)$ QCD. Through utilizing the parametrization of the $SU(2)$ gauge
potential and choosing proper gauge, they showed that the monopole potential
acquires a mass through the quantum correction after integrating out all the
dynamical degrees of the non-Abelian potential.

In this paper, from the similar consideration as that of Cho {\it et al}, we
obtain the effective theory of $SU(N)$ QCD based on the first principles. It
is found that there exists mass gaps in the infra-red limit of $SU(N)$ QCD
corresponding to different Abelian projection parts of $SU(N)$ connection.
Our result gives a connection between the generalized non-linear model of
Skyrme-Faddeev type and $SU(N)$ QCD in the low energy limit. We also shows
this effective theory may have special feature at large $N$ limit and may
relate to the properties of $D$-brane. In order to achieve the goal desired,
we also give a general decomposition of $SU(N)$ connection in terms of $N-1$
orthonormal vectors defined by conjugating the Cartan matrices by a generic $%
SU(N)$ element. The $SU(N)$ connection we give corresponds to Cho connection
and Faddeev-Niemi connection under different condition.

This paper is arranged as follows. In section 2, we give the general
decomposition of $SU(N)$ connection. In section 3, the effective theory of $%
SU(N)$ QCD is given. At last, from the viewpoint of background field, we
give a effective lagrangian of Skyrme-Faddeev type at section 4.

\section{Decomposition of $SU(N)$ gauge potential}

Let ${\bf M}$ be a $n$-dimensional Riemannian manifolds and $P(\pi ,{\bf M}%
,G)$ be a principal bundle with the structure group $G=SU(N)$. A smooth
vector field $\phi ^a$ $(a=1,2,...,N^2-1)$ can be found on the base manifold 
${\bf M}$ (a section of a vector bundle over ${\bf M}${\bf ). }The covariant
derivative 1-form of $\phi ^a$ is 
\begin{equation}
D\phi ^a=d\phi ^a+f^{abc}A^b\phi ^c
\end{equation}
where $A^a$ is the connection 1-form 
\begin{equation}
A^a=A_\mu ^adx^\mu .
\end{equation}
The curvature 2-form is defined as 
\begin{equation}
F^a=\frac 12F_{\mu \nu }^adx^\mu \wedge dx^\nu =dA^a+gf^{abc}A^b\wedge A^c
\end{equation}
The defining representation of the $SU(N)$ Lie algebra consists of $N^2-1$
matrices $T^a$ which satisfy 
\begin{equation}
T^aT^b=\frac 1{2N}\delta ^{ab}+\frac i2f^{abc}T^c+\frac 12d^{abc}T^c,
\end{equation}
where $f^{abc}$ are completely antisymmetric and $d^{abc}$ are completely
symmetric. In terms of $T^a$, the vector field $\phi ^a$ can be represented
as matrix form 
\begin{equation}
\phi =\phi ^aT^a
\end{equation}
and its covariant derivative as 
\begin{equation}
D\phi =d\phi -ig[A,\phi ]
\end{equation}
Conjugating the Cartan matrices $H_i$ $(i=1,2,...,N-1)$ by a generic element 
$g\in SU(N),$ we define $N-1$ vectors 
\begin{equation}
m_i=m_i^aT^a=gH_ig^{-1}
\end{equation}
which are othornormal to each other 
\begin{equation}
\frac 12Tr(m_im_j)=m_i^am_j^a=\delta _{ij}.
\end{equation}
In this way, $m_i$ make up an over determined set of coordinates on the
orbit $SU(N)/U(1)^{N-1}$. The covariant derivative of $m_i^a$ are 
\begin{equation}
D_\mu m_i^a=dm_i^a+gf^{abc}A_\mu ^bm_i^c  \label{cov-m}
\end{equation}
By making use of a relation 
\begin{equation}
f^{acd}f^{bed}m_i^cm_i^e=\delta ^{ab}-m_i^am_i^b
\end{equation}
it is easy to solve the equation (\ref{cov-m}) to get gauge potential $A$
expressed in terms of $m_i$ as 
\begin{equation}
A_\mu =\frac 1gf^{abc}\partial _\mu m_i^bm_i^c-\frac 1gf^{abc}D_\mu
m_i^bm_i^c+A_{i\mu }m_i^a  \label{decomposition}
\end{equation}
in which $A_{i\mu }$ is the projection of $A_\mu $ on the $m_i$%
\begin{equation}
A_{i\mu }=A_\mu ^am_i^a
\end{equation}
If $m_i^a$ are covariant constants 
\begin{equation}
D_\mu m_i^a=0
\end{equation}
we get the Cho connection\cite{cho1} 
\begin{equation}
A_\mu =\frac 1gf^{abc}\partial _\mu m_i^bm_i^c+A_{i\mu }m_i^a
\end{equation}
On 4-dimensional manifold, by choosing a complete basis of subspace of $%
su(N) $ Lie algebra which is orthogonal to the subspace make up by $m_i$,
the covariant part of the gauge potential (\ref{decomposition}) can be
expressed as 
\begin{equation}
-f^{abc}D_\mu m_i^bm_i^c=\rho ^{ij}f^{abc}\partial _\mu m_i^bm_j^c+\sigma
^{ij}d^{abc}\partial _\mu m_i^bm_j^c
\end{equation}
and the connection is expressed as 
\begin{equation}
A_\mu =A_{i\mu }m_i^a+\frac 1gf^{abc}\partial _\mu m_i^bm_i^c+\frac 1g\rho
^{ij}f^{abc}\partial _\mu m_i^bm_j^c+\frac 1g\sigma ^{ij}d^{abc}\partial
_\mu m_i^bm_j^c
\end{equation}
which is just the connection proposed by Faddeev and Niemi\cite{fadeev}.

Define a covariant vector 1-form $X^a=X_\mu ^adx^\mu $ as the covariant part
of the connection (\ref{decomposition}) 
\begin{equation}
X^a=-\frac 1gf^{abc}Dm_i^bm_i^c
\end{equation}
which is orthogonal to $m_i^a$%
\begin{equation}
X^am_i^a=0
\end{equation}
and denote the non-covariant part of (\ref{decomposition}) as 
\begin{equation}
\hat{A}_\mu ^a=A_{i\mu }m_i^a+C_\mu ^a
\end{equation}
where 
\begin{equation}
C_\mu ^a=\frac 1gf^{abc}\partial _\mu m_i^bm_i^c
\end{equation}
We can rewrite the gauge potential simply as 
\begin{equation}
A_\mu ^a=\hat{A}_\mu ^a+X_\mu ^a
\end{equation}
The connection we obtained naturally pick out the Abelian parts from the
full connection. In the expression (\ref{decomposition}) $A_{i\mu }$ serve
as ''electric'' potential and $C_\mu $ serve as ''magnetic'' potential.
Under the infinitesimal gauge transformation 
\begin{equation}
\delta m_i^a=-f^{abc}\alpha ^bm_i^c,\quad \delta A_\mu ^a=\frac 1gD_\mu
\alpha ^a
\end{equation}
one has 
\begin{equation}
\delta A_{i\mu }=\frac 1gm_i^a\partial _\mu \alpha ^a,\quad \delta \hat{A}%
_\mu ^a=\frac 1g\hat{D}_\mu \alpha ^a,\quad \delta X_\mu ^a=-f^{abc}\alpha
^bX_\mu ^c
\end{equation}
Using the $SU(N)$ connection obtained above, the gauge field can be
expressed as 
\begin{eqnarray}
F_{\mu \nu }^a &=&m_i^a(\partial _\mu A_{i\nu }-\partial _\nu A_{i\mu })-%
\frac 1gm_i^af^{bcd}m_i^b\partial _\mu m_j^c\partial _\nu m_j^d  \nonumber \\
&&+\hat{D}_\mu X_\nu ^a-\hat{D}_\nu X_\mu ^a+gf^{abc}X_\mu ^bX_\nu ^c
\end{eqnarray}
in which 
\begin{equation}
\hat{D}_\mu X_\nu ^a=\partial _\mu X_\nu ^a+gf^{abc}\hat{A}_\mu ^bX_\nu ^c
\end{equation}

\section{Effective theory of $SU(N)$ QCD}

The Yang-Mills Lagrangian can be expressed in terms of $A_{i\mu }$ and $%
X_\mu ^a$%
\begin{eqnarray}
{\cal L} &=&{\cal -}\frac 14(F_{\mu \nu }^a)^2  \nonumber \\
&=&-\frac 14\hat{F}_{i\mu \nu }^2-\frac g2f^{abc}\hat{F}_{\mu \nu }^aX^{b\mu
}X^{c\nu }-\frac{g^2}4(f^{abc}X_\mu ^bX_\nu ^c)^2-\frac 14(\hat{D}_\mu X_\nu
^a-\hat{D}_\nu X_\mu ^a)^2  \label{lagrangian}
\end{eqnarray}
where $\hat{F}_{i\mu \nu }$ is defined as 
\begin{equation}
\hat{F}_{i\mu \nu }=\partial _\mu A_{i\nu }-\partial _\nu A_{i\mu }-\frac 1g%
f^{bcd}m_i^b\partial _\mu m_j^c\partial _\nu m_j^d
\end{equation}
To get (\ref{lagrangian}), we assume 
\begin{equation}
f^{abc}X_\mu ^bX_\nu ^c=m_i^af^{bcd}m_i^bX_\mu ^cX_\nu ^d  \label{assum}
\end{equation}
which means the Lie product of $X_\mu $ is belong to the subspace made up by 
$m_i$. This condition is satisfied intrinsically at $SU(2)$ gauge theory.
The further analysis of this assumption will be showed in the forth coming
paper.

In order to get the effective action expressed solely by the topological
fields $m_i$, we need to integrate out all the dynamical degrees. Consider
the generating function for (\ref{lagrangian}) 
\begin{equation}
W[J_{i\mu },J_\mu ^a]=\int DA_{i\mu }DX_\mu ^a\exp [-i\int (\frac 14F_{\mu
\nu }^aF^{a\mu \nu }+A_{i\mu }J_{i\mu }+X_\mu ^aJ_\mu ^a)d^4x].
\end{equation}
To integrate out all the dynamical degrees we need to choose a proper gauge
to leave $m_i$ as a background. The techniques we use here are similar to
that were used by Y. M.. Cho et al\cite{cho2}.

Firstly, we fix $m_i$ as a background such that under the infinitesimal
gauge transformation 
\begin{equation}
\delta m_i^a=0,\quad \delta A_\mu ^a=\frac 1gD_\mu \alpha ^a
\end{equation}
from which one has 
\begin{equation}
\delta A_{i\mu }=\frac 1gm_i^aD_\mu \alpha ^a,\quad \delta \hat{A}_\mu ^a=%
\frac 1g(m_i^bD_\mu \alpha ^b)m_i^a,\quad \delta X_\mu ^a=\frac 1g(D_\mu
\alpha ^a-m_i^am_i^bD_\mu \alpha ^b)
\end{equation}
Fix the gauge with the condition 
\begin{eqnarray}
F_1^a &=&\hat{D}_\mu X_\mu ^a-f_1=0 \\
{\cal L}_{gf_1} &=&-\frac 1{2\kappa }(\hat{D}_\mu X_\mu ^a)^2
\end{eqnarray}
The corresponding Faddeev-Popov determinant can be written as 
\begin{equation}
M_1^{ab}=\frac{\delta F_1^a}{\delta \alpha ^b}=\frac 1g(\hat{D}_\mu D_\mu
)^{ab}
\end{equation}
With this the generating functional takes the form with $\kappa =1,$%
\begin{eqnarray}
W[J_\mu ^a] &=&\int DX_\mu ^a\det ||M_1||\exp \{-i\int [\frac 14\hat{F}_{\mu
\nu }^2-\frac 12X_\nu ^a\hat{D}_\mu \hat{D}_\mu X_\nu ^a+gf^{abc}\hat{F}%
_{\mu \nu }^aX_\mu ^bX_\nu ^c  \nonumber \\
&&+\frac{g^2}4(f^{abc}X_\mu ^bX_\nu ^c)^2+X_\mu ^aJ_\mu ^a]d^4x\}
\end{eqnarray}
where $\hat{F}_{\mu \nu }^a$ is defined as 
\begin{equation}
\hat{F}_{\mu \nu }^a=m_i^a(\partial _\mu A_{i\nu }-\partial _\nu A_{i\mu })-%
\frac 1gm_i^af^{bcd}m_i^b\partial _\mu m_j^c\partial _\nu m_j^d
\end{equation}
To remove the quartic term of $X_\mu ^a$, we introduce $N^2-1$ auxiliary
antisymmetric tensor field $\chi _{\mu \nu }^a$ and express the quartic term
as 
\begin{equation}
\exp [-\frac i4\int (f^{abc}X_\mu ^bX_\nu ^c)^2d^4x]=\int D\chi _{\mu \nu
}^a\exp [-\frac i4\int (\chi _{\mu \nu }^a\chi _{\mu \nu }^a-2i\chi _{\mu
\nu }^af^{abc}X_\mu ^bX_\nu ^c)d^4x]
\end{equation}
The integration over $X_\mu ^a$ results in the functional determinant 
\begin{equation}
\det\nolimits^{-\frac 12}K_{\mu \nu }^{ab}=\det\nolimits^{-\frac 12}[g_{\mu
\nu }\delta ^{ab}\hat{D}\hat{D}-2gf^{abc}\hat{F}_{\mu \nu }^c-if^{abc}\chi
_{\mu \nu }^c]
\end{equation}
so that the generating function is given by 
\begin{equation}
W[J_\mu ^a]=\int D\chi _{\mu \nu }^a\det ||M_1||\det\nolimits^{-\frac 12%
}||K||\exp \{-i\int [\frac 14\hat{F}_{\mu \nu }^2+\frac 14(\chi _{\mu \nu
}^a)^2+\frac 12J_\mu ^aK_{\mu \nu }^{-1}J_\nu ^a]d^4x
\end{equation}
Take the trivial classical configurations $\chi _{\mu \nu }^a=0$, we can
perform the integrate over the auxiliary antisymmetric field. In one loop
approximation, by making use of the dimensional regulation, we can calculate
the determinants and those terms involving only divergent parts are 
\begin{eqnarray}
\ln \det ||M_1|| &=&i\frac{Ng^2}{96\pi ^2}\frac 1\varepsilon \int \hat{F}%
_{\mu \nu }^2d^4x  \nonumber \\
\ln \det\nolimits^{-\frac 12}||K|| &=&i\frac{10Ng^2}{96\pi ^2}\frac 1%
\varepsilon \int \hat{F}_{\mu \nu }^2d^4x
\end{eqnarray}
So the effective Lagrangian for the restricted QCD is written as 
\begin{equation}
{\cal L}_{(R)eff}=-(\frac 14-\frac{11Ng^2}{96\pi ^2}\frac 1\varepsilon )\hat{%
F}_{\mu \nu }^2
\end{equation}
This restricted QCD action contains only the Abelian projection which
nevertheless has the full non-Abelian gauge degrees of freedom.

Then we consider 
\begin{equation}
W[J_{i\mu }]=\int DA_{i\mu }\exp \{-i\int [(\frac 14-\frac{11Ng^2}{96\pi ^2}%
\frac 1\varepsilon )\hat{F}_{\mu \nu }^2+A_{i\mu }J_{i\mu }]d^4x  \label{w2}
\end{equation}
We have to integrate out the electric degrees $A_{i\mu }$ from the
restricted QCD to obtain the effective action for $m_i$. We can choose the
gauge condition as 
\begin{eqnarray}
F_2^a &=&\partial _\mu A_{\mu i}m_i^a+\partial _\mu C_\mu ^a-f_2  \nonumber
\\
&=&\partial _\mu A_{\mu i}m_i^a+\frac 1gf^{abc}\partial ^2m_i^bm_i^c-f_2=0
\label{f2} \\
{\cal L}_{gf_2} &=&-\frac 1{2\lambda }[(\partial _\mu A_{\mu i})^2+(\partial
_\mu C_\mu ^a)^2]  \label{g2}
\end{eqnarray}
One can easy to obtain the Faddeev-Popov determinant as 
\begin{eqnarray}
M_2^{ab} &=&\partial ^2\delta ^{ab}+(m_i^a\partial _\mu
m_i^b-2f^{aec}f^{cdb}m_i^e\partial _\mu m_i^d)\partial _\mu  \nonumber \\
&&+f^{aec}f^{cdb}(\partial ^2m_i^em_i^d-m_i^e\partial
^2m_i^d)-f^{abc}\partial _\mu A_{\mu i}m_i^c
\end{eqnarray}
Calculate the determinant in one-loop approximation, one can get 
\begin{eqnarray}
\ln \det ||M_2|| &=&i\int -\frac{N+2}{32\pi ^2\varepsilon }\mu
_0^2g^2(C_{i\mu }^a)^2+\frac 1{192\pi ^2\varepsilon }(\frac N4g^2(\partial
_\mu C_\nu ^a)^2-\frac 32NTr\{C_\mu C_\mu C_\nu C_\nu \}  \nonumber \\
&&+(\frac{11}4N-3)g^2(\partial _\mu C_\mu ^a)^2+(6+\frac{5N}8%
)g^2(m_k^a\partial _\nu C_\mu ^a)^2  \nonumber \\
&&-3\partial ^2m_i^a\partial ^2m_i^a-7\partial _\mu m_i^b\partial _\nu
m_j^b\partial _\mu m_i^d\partial _\nu m_j^d-\frac 12\partial _\mu
m_i^b\partial _\mu m_j^b\partial _\nu m_i^d\partial _\nu m_j^d  \nonumber \\
&&-\frac 38\partial _\mu m_i^b\partial _\mu m_i^b\partial _\nu m_j^d\partial
_\nu m_j^d-\frac 34\partial _\mu m_i^b\partial _\nu m_i^b\partial _\mu
m_j^d\partial _\nu m_j^d)  \label{lnm2}
\end{eqnarray}
where $\mu _0$ is a mass scale. Integrate $A_\mu $ from (\ref{w2}) with (\ref
{g2}) and (\ref{lnm2}), with $\lambda =1$. The effective Lagrangian is
obtained 
\begin{eqnarray}
{\cal L}_{eff} &=&-\frac{N+2}{32\pi ^2}\mu _0^2\frac 1\varepsilon (\partial
_\mu m_i^a)^2-\frac 14(1-\frac 1{8\pi ^2}\frac 1\varepsilon -\frac{181}{384}%
\frac N{\pi ^2}\frac 1\varepsilon )g^2(m_k^a\partial _\nu C_\mu ^a)^2 
\nonumber \\
&&+(-\frac 1{2g^2}+\frac 1{192\pi ^2}(\frac{11}4N-3)\frac 1\varepsilon
)g^2(\partial _\mu C_\mu ^a)^2+\frac 1{768}\frac N{\pi ^2}\frac 1\varepsilon
g^2(\partial _\mu C_\nu ^a)^2  \nonumber \\
&&-\frac N{128\pi ^2}\frac 1\varepsilon g^4Tr(C_\mu C_\mu C_\nu C_\nu )-%
\frac 1{64\pi ^2}\frac 1\varepsilon \partial ^2m_i^a\partial ^2m_i^a-\frac 7{%
192\pi ^2}\frac 1\varepsilon (\partial _\mu m_i^b\partial _\nu m_j^b)^2 
\nonumber \\
&&-\frac 1{384\pi ^2}\frac 1\varepsilon (\partial _\mu m_i^b\partial _\mu
m_j^b)^2-\frac 1{384\pi ^2}\frac 1\varepsilon (\partial _\mu m_i^b\partial
_\nu m_i^b)^2-\frac 1{512\pi ^2}\frac 1\varepsilon (\partial _\mu
m_i^b\partial _\mu m_i^b)^2
\end{eqnarray}
$\ $After a proper renormalization the final effective Lagrangian can be
written as

\begin{eqnarray}
{\cal L}_{eff} &=&-\frac{\mu ^2}2(\partial _\mu m_i^a)^2-\frac 14%
(m_k^a\partial _\nu C_\mu ^a)^2-\alpha _1(\partial _\mu C_\mu ^a)^2-\alpha
_2(\partial _\mu C_\nu ^a)^2  \nonumber \\
&&-\alpha _3Tr(C_\mu C_\mu C_\nu C_\nu )-\alpha _4(\partial
^2m_i^a)^2-\alpha _5(\partial _\mu m_i^a\partial _\nu m_j^a)^2  \nonumber \\
&&-\alpha _7(\partial _\mu m_i^a\partial _\mu m_j^a)^2-\alpha _6(\partial
_\mu m_i^a\partial _\mu m_i^a)^2-\alpha _7(\partial _\mu m_i^a\partial _\nu
m_i^a)^2  \label{fin-lang}
\end{eqnarray}
where $\mu $ and $\alpha $ are the renormalized coupling constants. This is
nothing but a generalized Skyrme-Faddeev Lagrangian to $SU(N)$. One can find
it is much complex than the case of $SU(2)$ QCD. We have obtained the
effective action of $SU(N)$ QCD from the first principles.

\section{Skyrme-Faddeev-like effective lagrangian of $SU(N)$ QCD}

Actually, in the above procedure, we have integrated out $X_\mu ^a$ and $%
A_\mu $ separately in two steps to emphasize the importance of the
restricted theory QCD \cite{cho1}. But certainly we could integrate out them
simultaneously in one step to obtain the desired effective action. For
example, such a gauge 
\begin{equation}
F^a=\partial _\mu A_{i\mu }m_i^a+\hat{D}_\mu X_\mu ^a+\partial _\mu C_\mu
^a-f^a=0
\end{equation}
is a kind of suitable choice. In this case, the final result is basically
similar. Furthermore, to study our problem conveniently, seeking a suitable
gauge is very meaningful as will be shown in the following.

In our decomposition, one could regard $C_\mu ^a$ either as a covariant
multiplet or simply as a fixed background. The first point of view provides
an active type by (\ref{fin-lang}), but the second point of view gives the
following passive type.

Fix the gauge with the condition 
\begin{equation}
F^a=\tilde{D}_\mu (A_{i\mu }m_i^a-C_\mu ^a)-f^a=0
\end{equation}
which $\tilde{D}_\mu $ is defined only by $C_\mu ^a$. The Lagrangian of
ghost is 
\begin{equation}
{\em L}_{gf}=-\frac 1{2\xi }[(\partial _\mu A_\mu )^2+(\tilde{D}_\mu
X^a{}_\mu )^2]
\end{equation}
so with () the generating function is 
\begin{eqnarray}
W[J_\mu ,J_\mu ^a] &=&\int DA_\mu DX_\mu ^aDc^aDc^{*a}\exp \{i\int [-\frac 14%
(F_{\mu \nu }^a)^2+c^{*a}\tilde{D}_\mu D_\mu c^a-\frac 1{2\xi }(\partial
_\mu A_{i\mu })^2  \nonumber \\
&&-\frac 1{2\xi }(\tilde{D}_\mu X_\mu ^a)^2-A_{i\mu }J_{i\mu }-X_\mu ^aJ_\mu
^a]d^4x\},
\end{eqnarray}
where $c$ and $c^{*}$ are ghost fields. In one loop approximation, the $%
X_\mu ^a$ and the ghost fields integrations give the following functional
determinations (with $\xi =1$) 
\begin{eqnarray*}
\ln Det^{-\frac 12}K_{\mu \nu }^{ab} &\simeq &\ln Det^{-\frac 12}[g_{\mu \nu
}(\tilde{D}\tilde{D})^{ab}-2gH_{\mu \nu }f^{abc}m_i^c] \\
&\simeq &-\frac{i5g^2N}{48\pi ^2\varepsilon }\int d^4x(\vec{H}_{\mu \nu })^2
\end{eqnarray*}
\begin{equation}
\ln DetM_{FP}^{ab}\simeq \frac{ig^2N}{24\pi ^2\varepsilon }\int d^4x(\vec{H}%
_{\mu \nu })^2.
\end{equation}
In conclusion, the effective action is right the Skyrme-Faddeev type 
\begin{equation}
{\em L}_{eff}=-\frac{\mu _0^2g^2N}{8\varepsilon }(C_\mu ^a)^2-\frac N{16\pi
^2\varepsilon }(H_{\mu \nu }^a)^2,
\end{equation}
where 
\begin{equation}
H_{\mu \nu }^a=\partial _\mu C_\nu ^a-\partial _vC_\mu ^a+f^{abc}C_\mu
^bC_\nu ^c.
\end{equation}
After renormalization the final effective Lagrangian can be written as 
\[
{\em L}_{eff}=-\frac{\mu _0^2}2(C_\mu ^a)^2-\frac 14(H^a{}_{\mu \nu })^2 
\]
which is similar to the Lagrangian Faddeev and Niemi expect\cite{fadeev}.
This is just the Skyrme-Faddeev-like lagrangian which allows the topological
knot solutions. Our solutions shows there exist connection between the
generalized non-linear sigma model of Skyrme-Faddeev type and $SU(N)$ QCD.

In the above discussion, we construct the effective theory for $SU(N)$ QCD.
Our analysis shows, just like $SU(2)$ QCD, there exist the mass gaps
corresponding to $m_i$ in the infra-red limit of QCD based on the first
principles. The monopole condensation can serve as the physical mechanism
for the confinement of color in $SU(N)$ QCD. However, one can find the mass
gotten for different $m_i$ are the same. The reason is that we choose only
one gauge condition (\ref{f2}) for all $m_i$. Alternatively, one can choose $%
N$ gauge condition for every $m_j=n$ such as 
\begin{equation}
F_{2j}^a=\partial _\mu A_{\mu j}n^a+f^{abc}\partial ^2m_i^bm_i^c-f_2=0
\end{equation}
Then $N$ different mass $\mu _j$ are obtained and the mass term of the
effective Lagrangian becomes 
\begin{equation}
-\frac{\mu _j^2}2(\partial _\mu m_j^a)^2.
\end{equation}
When calculating the Yang-Mills Lagrangian, we assume (\ref{assum}) to
remove a cross term. This assumption may be superfluous and do not affect
the effective Lagrangian we obtain too much for it involves only the
covariant parts of the gauge potential.

The result gotten in section 3 is not the same as that gotten in section 4.
The reason is the different choices of gauge condition. We notice an
important feature of the former is that when $N\rightarrow \infty $, some
special terms of the Lagrangian (\ref{fin-lang}) manifest their importances
and the others can be ignored 
\begin{eqnarray}
{\cal L}_{eff} &=&\frac{\mu _0^2}2(\partial _\mu m_i^a)^2+\frac 14%
(m_k^a\partial _\nu C_\mu ^a)^2  \nonumber \\
&&+\alpha _2(\partial _\mu C_\mu ^a)^2+\alpha _3(\partial _\mu C_\nu
^a)^2-\alpha _4Tr(C_\mu C_\mu C_\nu C_\nu \}
\end{eqnarray}
We still don't know what problem it will cause. But we think it is crucial
at the large $N$ cases, and may relate to some properties of $D$-brane. And
we believe the last term in above Lagrangian will give rise to many
interesting physical effects.

\end{document}